\documentclass[reprint]{revtex4-1}
\usepackage{amsmath}
\usepackage{amssymb}
\usepackage{graphicx}
\usepackage{bm}
\usepackage{natbib}

\begin{document}
\title{Quantifying Gyrotropy in Magnetic Reconnection}

\author{M. Swisdak}
\email{swisdak@umd.edu}


\begin{abstract}

A new scalar measure of the gyrotropy of a pressure tensor is
defined. Previously suggested measures are shown to be incomplete by
means of examples for which they give unphysical results.  To
demonstrate its usefulness as an indicator of magnetic topology, the
new measure is calculated for electron data taken from numerical
simulations of magnetic reconnection, shown to peak at separatrices
and X-points, and compared to the other measures.  The new diagnostic
has potential uses in analyzing spacecraft observations and so a
method for calculating it from measurements performed in an arbitrary
coordinate system is derived.

\end{abstract}
\maketitle


\section{Introduction}

The term magnetic reconnection, as commonly understood, refers to a
process whereby the energy of the embedded field is transferred to the
surrounding plasma via bulk acceleration and Ohmic dissipation.  While
frequently associated with changes in the topology of the magnetic
field, precisely defining where and how reconnection occurs is a
surprisingly subtle task \citep{schindler88a}.

In what are known as 2.5D geometries, where variations in one
direction are suppressed, the coupling of the reduced dimensionality
with the divergence-free nature of the magnetic field allows field
lines to be completely characterized as contours of a scalar function,
$\psi$.  ($\psi$ is equivalent to the component of the magnetic vector
potential in the invariant direction.)  Saddle points of $\psi$, known
as X-points, mark locations of both topological change and magnetic
reconnection.  In fully three-dimensional systems, field lines can no
longer be described by a single function and characterizing
reconnection becomes significantly more difficult.  Finding the
locations where magnetic reconnection occurs is non-trivial, even in
numerical simulations \citep{dorelli08a,haynes10a,komar13a}.

Yet simulations have the luxury of a synoptic view with, in principle,
access to the complete description of the plasma at any point in the
domain.  Spacecraft observations, in contrast, must make due with a
limited set of measurements available from, at most, a few locations.
The topological identification of reconnection sites is non-local in
nature and hence extremely difficult to implement with these
restrictions.  Determinations must instead be based on local
measurements of the electromagnetic fields and particles.

Due to their small Larmor radii and brief gyroperiods, electrons are
closely tied to the magnetic field, much more so than the heavier
ions.  By streaming for long distances along field lines they
efficiently probe magnetic structure and so locations of topological
interest, such as reconnection sites, should leave signatures in
electron measurements.  \cite{vasyliunas75a} identified one such
signature by noting that, because of finite Larmor radius effects, the
guiding center approximation breaks down near an X-point.  One
consequence is that reconnection, at least in 2.5D, requires an
asymmetry (specifically, non-gyrotropy) in the electron pressure
tensor there.  Spacecraft measurements of electron distribution
functions can hence be used as a proxy for topology.  In this paper we
present a diagnostic quantifying the size of departures from gyrotropy
and demonstrate, through numerical simulations, that it faithfully
maps regions associated with magnetic reconnection.

\section{Measuring Gyrotropy}\label{measure}

Given a distribution function $f(\mathbf{x},\mathbf{v})$ that
describes the density of particles with mass $m$ in a phase space
defined by position ($\mathbf{x}$) and velocity ($\mathbf{v}$), the
pressure $\mathbb{P}$ can be defined as
\begin{equation}
\mathbb{P} = m \int \mathbf{w}\mathbf{w} f\,\mathrm{d}^3\mathbf{v}\text{ ,}
\end{equation}
where $\mathbf{w} = \mathbf{v} - \mathbf{\bar{v}}$ is the velocity
relative to the mean flow $\mathbf{\bar{v}}$.  While $\mathbb{P}$ is,
in general, a tensor with six independent components, it can be
characterized by a single scalar when $f$ is spherically symmetric
(e.g., a Maxwellian).  A magnetic field, by imposing a preferred
direction, can produce a cylindrically symmetric, or gyrotropic,
distribution.  In this case the pressure tensor takes the form
\begin{equation}\label{algebra}
\mathbb{P} = P_{\parallel}\widehat{\mathbf{b}}\widehat{\mathbf{b}} +
P_{\perp}(\mathbb{I} - \widehat{\mathbf{b}}\widehat{\mathbf{b}})\text{
  ,}
\end{equation}
where $\widehat{\mathbf{b}} = |\mathbf{B}|/B$ is a unit vector in the
direction of the magnetic field, $P_{\parallel}$ and $P_{\perp}$ are
scalars, and $\mathbb{I}$ is the unit tensor.  In the equivalent
matrix formulation,
\begin{equation}\label{gyromat}
\mathbb{P} = \left( \begin{array}{ccc}
P_{\parallel} & 0 & 0 \\
0 & P_{\perp} & 0 \\
0 & 0 & P_{\perp} \end{array} \right)\text{.}
\end{equation}
Observationally, pressure tensors are never completely gyrotropic
because of the presence of non-zero off-diagonal components, although
these are often much smaller than the diagonal terms.  Since
departures from gyrotropy are expected to be associated with locations
of magnetic reconnection, a quantitative measure of this smallness
would be useful.

With this goal in mind, \cite{scudder08a} proposed a measure, dubbed
the agyrotropy and denoted $A\varnothing_e$, that characterizes a
distribution function's weighted average of dispersions of velocities
perpendicular to the local field direction (see Appendix A of their
paper for a full description).  It varies between $0$ and $2$ with the
minimum supposedly corresponding to gyrotropic distributions.
However, as demonstrated below, there exist agyrotropic pressure
tensors for which $A\varnothing_e = 0$.

\cite{aunai13a} proposed another measure, denoted $D_{ng}$ and named
non-gyrotropy.  It is proportional to the root-mean-square of the
off-diagonal elements of the pressure tensor normalized by the local
thermal energy.  As with the agyrotropy, $D_{ng}=0$ for a gyrotropic
distribution but, as also shown below, there exist pressure tensors
with maximal departures from gyrotropy for which $D_{ng}$ is
arbitrarily close to 0.

A brief mathematical digression is necessary in order to justify a
better measure.  Because pressure tensors originate as moments of
distribution functions, their matrix representations must satisfy
certain constraints.  Specifically, a physically meaningful pressure
tensor is a real $3 \times 3$ symmetric matrix (i.e., $P_{ij} =
P_{ji}$) with non-negative eigenvalues.  Matrices possessing these
properties, including all pressure tensors, are called symmetric
positive semi-definite. If all of the eigenvalues are positive the
matrix is positive definite.  (That the eigenvalues of a pressure
tensor must be non-negative follows from the requirement that after
diagonalization, which is always possible for a real symmetric matrix,
the matrix elements are both the eigenvalues and the components of the
pressure along the basis vectors of the rotated frame.  For physical
distributions the latter, and hence the former, must be greater than
or equal to zero.)

Defining a scalar measure of gyrotropy requires a result from the
theory of symmetric positive semi-definite matrices.  It can be shown
\citep{horn12a} that a matrix is positive semi-definite if and only
if all principal minors -- the determinants of every submatrix arising
from deleting the same set of rows and columns -- are non-negative.
(If zero eigenvalues are excluded, so that the matrix is positive
definite, Sylvester's criterion applies \citep{horn12a}.  In this
case, a necessary and sufficient condition is that the leading
principal minors, i.e., the determinants of all upper-left
submatrices, be positive.)

A pressure tensor with gyrotropic diagonal, but non-zero off-diagonal,
entries can be written as
\begin{equation}\label{gyroform}
\mathbb{P} = \left( \begin{array}{ccc}
P_{\parallel} & P_{12} & P_{13} \\
P_{12} & P_{\perp} & P_{23} \\
P_{13} & P_{23} & P_{\perp} \end{array} \right)\text{.}
\end{equation}
In this representation one of the coordinate axes points in the
direction of the local magnetic field and the others are oriented such
that the final two components of the diagonal of $\mathbb{P}$ are
equal. (That this is always possible is demonstrated in the Appendix.)
Since $\mathbb{P}$ is positive semi-definite the result discussed
above implies the inequalities
\begin{equation}\label{syl}
P_{12}^2 \leq P_{\parallel}P_{\perp} \qquad P_{13}^2 \leq P_{\parallel}P_{\perp} \qquad 
P_{23}^2 \leq P_{\perp}^2\text{.}
\end{equation}
The converse does not hold: satisfaction of these inequalities does
not guarantee that $\mathbb{P}$ is positive semi-definite.

Adding the inequalities implies a natural definition for a measure of
gyrotropy:
\begin{equation}\label{agyro}
Q=\frac{P_{12}^2 + P_{13}^2 + P_{23}^2}{P_{\perp}^2 +
  2P_{\perp}P_{\parallel}}\text{ .}
\end{equation}
For gyrotropic tensors $Q=0$, while for maximal departures from
gyrotropy $Q=1$.  Evaluating $Q$ from equation \ref{agyro} requires
$\mathbb{P}$ to be in the form given by equation \ref{gyroform} even
though, in general, $\mathbb{P}$ is measured in an arbitrary
coordinate system.  While it is always possible to rotate $\mathbb{P}$
into a frame in which it has the form of equation \ref{gyroform}, the
Appendix demonstrates how to calculate $Q$ without such a
transformation by using tensor invariants.

The theory of positive semi-definite matrices provides a firm
mathematical basis for the definition of $Q$.  With it in mind, it is
possible to construct pressure tensors for which other proposed
measures give physically unreasonable results.  Consider a system with
a pressure tensor, of the type shown in equation \ref{gyroform}, given
by
\begin{equation}\label{sc_ex}
\mathbb{P} = \left( \begin{array}{ccc}
1 & 1/2 & 0 \\
1/2 & 1 & 0 \\
0 & 0 & 1 \end{array} \right)\text{.}
\end{equation}
Since $\mathbb{P}$ has eigenvalues of $1/2$, $1$, and $3/2$, it is
positive semi-definite and hence physically valid.  Yet, despite the
off-diagonal elements that make it clearly not gyrotropic,
\begin{equation}
A\varnothing_e = 0\text{ .}
\end{equation}
For reference, $\sqrt{Q} = \sqrt{3}/6$ and $D_{ng}=\sqrt{2}/3$ in this
case.  (The other gyrotropy measures considered here scale as ratios
of first powers of pressure components while $Q$ scales quadratically.
The subsequent discussion, where necessary to ensure an unbiased
comparison, uses $\sqrt{Q}$.)

In fact $A\varnothing_e=0$ for any positive semi-definite $\mathbb{P}$
in the form of equation \ref{gyroform} with $P_{23} = 0$, irrespective
of the other off-diagonal components.  This occurs because, by
construction, $A\varnothing_e$ only measures departures from symmetry
in the plane perpendicular to the magnetic field axis.  However, as
equation \ref{sc_ex} demonstrates, there are other ways in which
pressure tensors can depart from gyrotropy.  The vanishing of
$A\varnothing_e$ for such cases indicates that it does not fully
describe a pressure tensor's gyrotropy.

The nongyrotropy measure proposed by \cite{aunai13a} is defined as
\begin{equation}
D_{ng} = \frac{\sqrt{8(P_{12}^2+P_{13}^2+P_{23}^2)}}{P_{\parallel} +
  2P_{\perp}}\text{ .}
\end{equation}
The most significant difference between it and $\sqrt{Q}$ occurs in
the normalization, which for $D_{ng}$ is to the thermal energy,
$P_{\perp} + P_{\parallel}/2$.  As with the other measures, $D_{ng}=0$
for gyrotropic tensors.  The maximum value of $D_{ng}$ varies with
$P_{\parallel}/P_{\perp}$, which is not the case for either $Q$ or
$A\varnothing$.  Its global maximum of $\sqrt{8/3}$ occurs when
$P_{\parallel} = P_{\perp}$.

The implications of this variation become apparent through
consideration of the pressure tensor
\begin{equation}\label{au_ex}
\mathbb{P} = \left( \begin{array}{ccc}
x & \sqrt{x} & \sqrt{x} \\
\sqrt{x} & 1 & 1 \\
\sqrt{x} & 1 & 1 \end{array} \right)
\end{equation}
with $x\geq 0$. Eigenvalues of $0$, $0$, and $x+2$ mean that $\mathbb{P}$
is positive semi-definite.  The gyrotropy measures are
\begin{equation}\label{au_D}
D_{ng} = \frac{\sqrt{8(2x+1)}}{x+2}\text{,} \qquad \sqrt{Q} = 1\text{,}
\qquad \text{and }A\varnothing_e = 2\text{ .}
\end{equation}
This pressure tensor is maximally non-gyrotropic for any value of $x$
because any perturbation that increases the off-diagonal components
will drive one of the eigenvalues negative.  Any measure of the
gyrotropy for this $\mathbb{P}$ should thus be independent of $x$ and
should take its maximal value.  Yet $D_{ng}$ varies with $x$.  This
behavior does not depend on $\mathbb{P}$ having zero eigenvalues,
although such a case is the simplest to analyze.  Similar pressure
tensors exist that both have three positive eigenvalues exist and
approach arbitrarily close to maximal non-gyrotropy yet for which
$D_{ng}$ varies in a manner similar to Equation \ref{au_D}.

The examples of equations \ref{sc_ex} and \ref{au_ex} are instances
where the deficiencies of $A\varnothing_e$ and $D_{ng}$ in quantifying
gyrotropy are particularly clear.  To a lesser or greater degree
similar issues will be present for other pressure tensors, although in
any specific example it may be difficult to tease out the degree to
which the gyrotropy is mischaracterized.

\section{Simulations}\label{sims}

While the arguments of the previous section suggest that $Q$ is a good
measure of gyrotropy, it is a separate question as to whether it is
also useful as a proxy for the identification of reconnection sites.
As a first attempt at an answer, we investigate particle-in-cell
simulations performed with the code {\tt p3d} \citep{zeiler02a}.  It
employs units based on a field strength $B_0$ and density $n_0$, with
lengths normalized to the ion inertial length $d_i =c/\omega_{pi}$,
where $\omega_{pi}$ is the ion plasma frequency, and times to the ion
cyclotron time $\Omega_{i0}^{-1}$.

\subsection{Run 1: Anti-Parallel Reconnection}\label{gemlike}

We first consider a variation of the standard GEM Challenge
\citep{birn01a}.  The simulation domain has dimensions $(L_x,L_y) =
(51.2,25.6)$ with an initial magnetic field of $B_x = \tanh(y/w_0)$
and $w_0 = 0.5$.  There is no initial field component in the $z$
direction.  To ensure pressure balance, the density $n = n_b +
\text{sech}^2(y/w_0)$, with $n_b = 0.2$, and the electron and ion
temperatures are initially isotropic with $T_e = 1/12$ and $T_i =
5/12$.  The ion-to-electron mass ratio is 400 and the speed of light
is 40.  The spatial grid has resolution $\Delta x = 1/160$, which
means there are 8 gridpoints per electron inertial length ($d_e$) and
$\approx 3$ per electron Larmor radius in the maximum field.  There
are $100$ particles per cell for each species in the asymptotic
region.

\begin{figure}
\includegraphics[width=0.95\columnwidth]{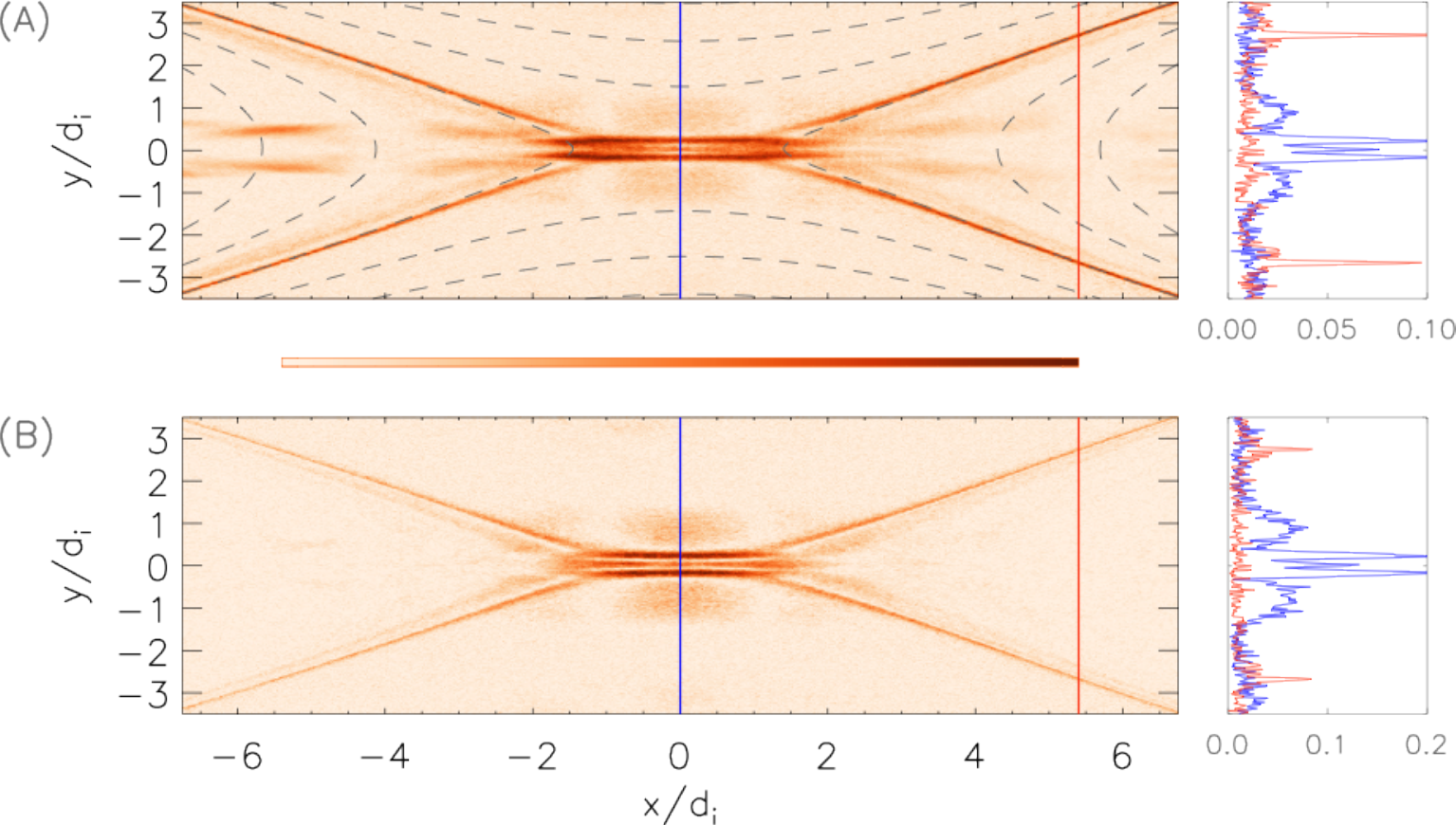}
\caption{\label{gyr1} Images of $\sqrt{Q}$ (panel A) and
  $A\varnothing_e/2$ (panel B) with superimposed magnetic field lines
  (dotted) for the run described in section \ref{gemlike}.  Colors
  represent different values in each image; the central bar shows the
  relative variation.  The data have been averaged over $0.03
  \Omega_{\text{ci}}^{-1} = 12 \Omega_{\text{ce}}^{-1}$, or $\approx
  2$ electron Larmor orbits in the asymptotic field, and over $0.5d_e$
  in each direction.  The panels on the right show cuts along the
  corresponding colored lines in the main images.  }
\end{figure}

Figure \ref{gyr1} shows a comparison of $\sqrt{Q}$ and
$A\varnothing_e/2$ during a time of steady reconnection ($D_{ng}$
closely follows $\sqrt{Q}$ in this case and so is not plotted).  The
magnetic field and pressure tensor components were averaged over
$\approx 2$ electron Larmor orbits and 4 gridpoints ($0.5d_e$) in each
direction before computing the gyrotropy measures.  Equation
\ref{qcalc} was used to calculate $\sqrt{Q}$; $A\varnothing_e$ was
calculated from the formulas given in Appendix A of \cite{scudder08a}.
In order to minimize subjectivity, no data manipulation (e.g., noise
rejection), other than the spatial and temporal averaging, was
performed.  Only a fraction of the simulation domain is pictured in
order to highlight the regions exhibiting strong departures from
gyrotropy.  On the right-hand side of each image are the traces of two
cuts, one through the X-point (blue) and the other $\approx 5d_i$
downstream (red).

Both $A\varnothing_e$ and $Q$ are significant near the X-point,
although there are minor differences in the aspect ratio and
structural details of the bright regions.  The bifurcation at $y/d_i =
0$ and $-2\lesssim x/d_i \lesssim 2$ mirrors that seen in the charge
and current densities (not shown) and reflects the complicated
dynamics of Speiser particle orbits near the region of null field.
The values within a few gridpoints of $(x,y)=(0,0)$ carry somewhat
more uncertainty because gyrotropy is an ill-defined quantity when
$B=0$.

While $\sqrt{Q}$ and $A\varnothing_e/2$ both vary between $0$ and $1$,
the more meaningful comparison is between the asymptotic value of each
cut, which gives a reasonable approximation of the inherent particle
noise, and the peak value at the X-point.  These are roughly equal for
the two measures.  On the other hand, the magnetic separatrices (red
peaks) are comparatively much weaker in $A\varnothing_e$ and stronger
in $\sqrt{Q}$.  The separatrices mark topological boundaries with, in
principle, zero width.  If the associated particle signatures exhibit
scales near the Larmor radius (as is known to be the case around the
X-line) then the intermingling between the upstream and downstream
plasmas should produce departures from gyrotropy there.

Both measures show humps just upstream of the X-point that indicate
small departures from gyrotropy.  These may arise from decreases in
$P_{\perp}$ (due to magnetic moment conservation) followed by
scattering to other components.  Another possibility is that beams of
inflowing electrons, when superimposed on the base Maxwellian, drive
the total distribution function away from cylindrical symmetry.

\subsection{Run 2: Guide-Field Reconnection}\label{ff}

As a second case we consider a force-free equilibrium where the
initial density ($n=1$) and temperatures ($T_i=T_e = 1/8$) have no
spatial variations.  The reconnecting field has the form $B_x =
\tanh(y/w_0)$, here with $w_0=1$, but now the out-of-plane (guide)
component $B_z$ varies so that $B_x^2+B_z^2$ is constant.  The initial
guide field is asymptotically $2$ and rises to $\sqrt{5}$ at the
current sheet's center.  The domain has dimensions $(L_x, L_y)= (51.2,
25.6)$, the ion-to-electron mass ratio is $25$, and the speed of light
is 15.  The spatial grid has resolution $\Delta x = 1/40$, which
implies 8 gridpoints per $d_e$ and $\approx 2$ per electron Larmor
radius in the maximum field.  There are $100$ particles per cell for
each species.

\begin{figure}
\includegraphics[width=0.95\columnwidth]{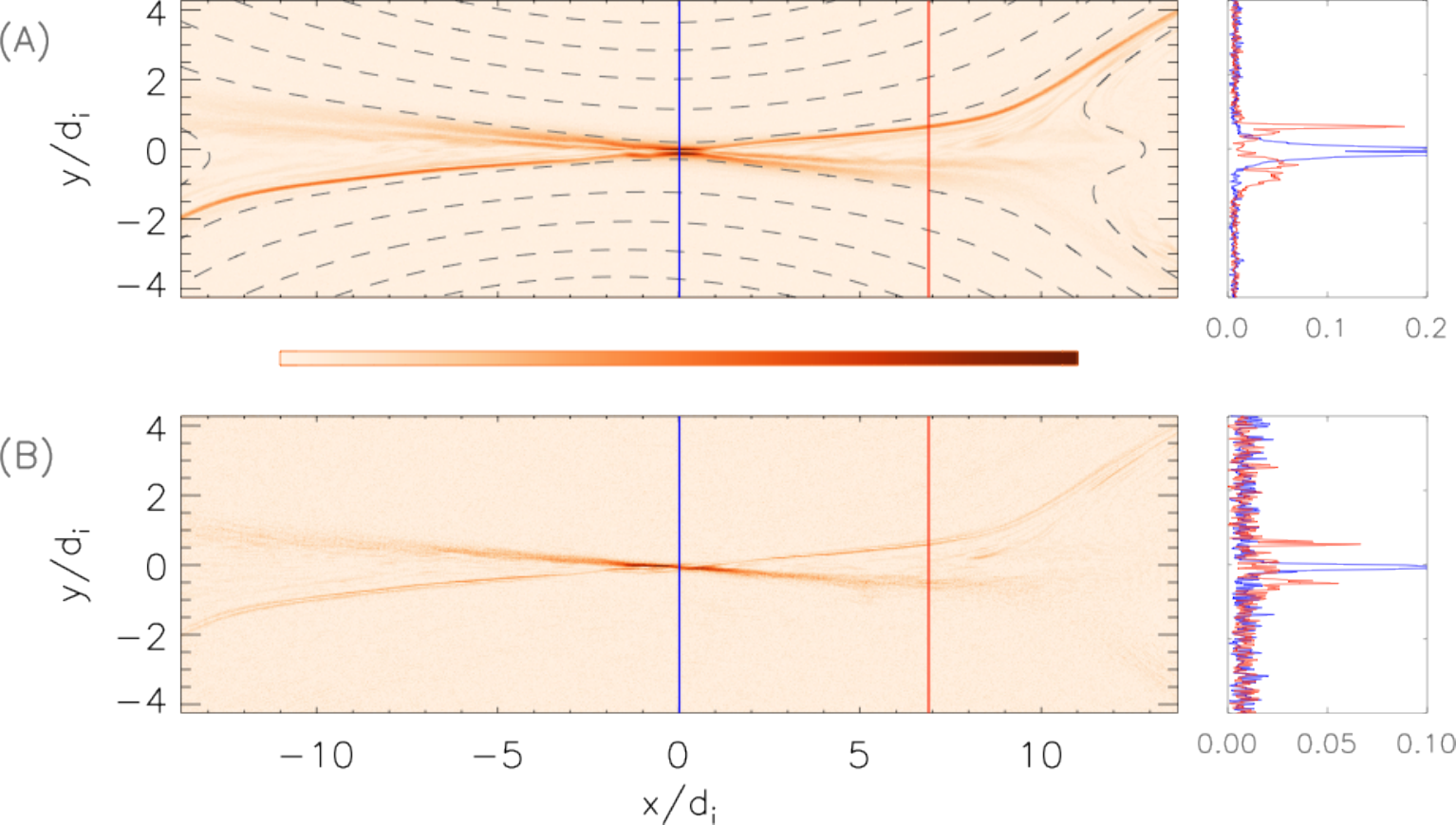}
\caption{\label{gyr2} Images of $\sqrt{Q}$ (panel A) and
  $A\varnothing_e/2$ (panel B), in the same format as Figure \ref{gyr1},
  for the run described in section \ref{ff}.  The data have been
  averaged over $0.5 \Omega_{\text{ci}}^{-1} = 12.5 \Omega_{\text{ce}}^{-1}$.}
\end{figure}

Figure \ref{gyr2}, in the same format as Figure \ref{gyr1}, shows the
region near the X-point.  Both $\sqrt{Q}$ and $A\varnothing_e$ peak
there, although again the morphologies in the downstream region
exhibit differences.  The asymmetry between the separatrices, a common
feature of guide-field reconnection, is apparent.  As in Figure
\ref{gyr1}, the cuts show that the separatrices are somewhat stronger
in $\sqrt{Q}$ than in $A\varnothing_e$.  In images of the entire
domain (not shown), the enhancement of $\sqrt{Q}$ on the upper-right
and lower-left separatrices continues around nearly the entire
exterior of the downstream magnetic island while $A\varnothing_e$
quickly peters out.  However, the strongest difference is in the
background noise level.  While the two measures have similar
peak-to-background ratios for the simulation of Section \ref{gemlike},
here they differ by a factor of $\approx 4$, with $\sqrt{Q}$ smoother
than in Figure \ref{gyr1} and $A\varnothing_e$ noisier.  The
improvement in $\sqrt{Q}$ can be attributed to the strong guide field
and its driving of the plasma towards gyrotropy.  The indentation in
the magnetic field lines at $x/d_i \approx 11$ is a transient feature
due to the interaction of the outflowing plasma and the remains of a
small magnetic island expelled from the X-point earlier in the
simulation.

\section{Discussion}\label{discussion}

We have proposed a new, mathematically rigorous measure of the
gyrotropy of an arbitrary pressure tensor and presented numerical
simulations of 2.5D reconnection demonstrating that gyrotropy
violations peak near magnetic topological boundaries.  Testing the
efficacy of $Q$ in three-dimensional simulations and with spacecraft
data are obvious next steps.  It is unlikely that any single
diagnostic can unfailingly and unambiguously identify reconnection
sites in spacecraft data, but $Q$ offers a complementary approach to
methods based on other measurements.  The calculation of $Q$ requires
the full electron pressure tensor -- available, for instance, on the
recently launched Magnetospheric Multiscale Mission -- and can be
easily performed in any coordinate system using the algorithm
presented in Appendix \ref{appendix}.

While $Q$ has some similarities to previously proposed measures,
particularly the non-gyrotropy parameter $D_{ng}$ of \cite{aunai13a},
there are pressure tensors (e.g., equation \ref{au_ex}) for which
$D_{ng}$ does not correctly characterize the gyrotropy.  However, in
many instances the two give similar results.  Their ratio can be
written as
\begin{equation}
\frac{Q}{D_{ng}^2} = \frac{(x+2)^2}{8(1+2x)} 
\end{equation}
where $x=P_{\parallel}/P_{\perp}$.  The right-hand side takes its
minimal value for $x=1$ and only varies by $\approx 30\%$ for $0\leq x
\leq 4$.  Although the ratio can become arbitrarily large when $x
\rightarrow \infty$, approaching that limit in a real system will
trigger other effects.  Of particular note is the firehose mode, for
which the linear instability criterion is
$\beta_{\perp}(P_{\parallel}/P_{\perp} -1) >2$.  Under normal
circumstances plasmas cannot significantly exceed this threshold,
which thus implies $x < 1+2/\beta_{\perp}$.  As a consequence, the
differences between $D_{ng}$ and $\sqrt{Q}$ are starkest for
$\beta_{\perp} \ll 1$.  (This argument is strictly only applicable for
gyrotropic plasmas because they are assumed in deriving the firehose
stability boundary.  It should be a reasonable approximation when the
pressure tensor is near gyrotropy, but significant off-diagonal
components will likely modify the instability threshold.)

The differences between $Q$ and the agyrotropy $A\varnothing_e$ are
more significant.  The latter only quantifies one type of departure
from gyrotropy, the breaking of cylindrical symmetry in the plane
perpendicular to the magnetic field, while the former handles the
general case.  Of more pressing interest to the interpretation of
simulations and observations, however, is how well the measures
identify regions of interesting magnetic topology.  Both successfully
illuminate X-lines for the simulations presented here, but $Q$ does a
better job of tracing magnetic separatrices, particularly in
guide-field reconnection.  Further analysis is necessary to make these
comparisons quantitative.

Finally, an interesting connection exists between $\mathbb{P}$ and the
inertia tensor of a rigid body.  Since both are symmetric and positive
semi-definite, a measure analogous to $Q$ also exists for the former.
To be relevant, however, something external to the body must define a
preferred direction in a manner similar to the role the magnetic field
plays in defining gyrotropy.

\begin{acknowledgments}
We would like to acknowledge helpful conversations with K. Schoeffler.
This work was supported by NASA grant NNX14AF42G.
\end{acknowledgments}

\appendix
\section{Calculating $Q$}\label{appendix}

Beginning with a pressure tensor in an arbitrary coordinate system,
the calculation of $Q$ from equation \ref{agyro} requires a
transformation into a frame in which the diagonal components are in
gyrotropic form.  A tensor $\mathbb{A}$ transforms under
coordinate rotations according to the prescription
\begin{equation}\label{rot}
\mathbb{A}^{\prime} = \mathbb{R}^T\mathbb{A}\mathbb{R}
\end{equation}
where $\mathbb{R}$ is a rotation matrix.  Both the symmetry of
$\mathbb{A}$ and its eigenvalues are preserved after rotation of the
coordinate axes. 

As an example, we demonstrate that, beginning with a pressure tensor
of the form
\begin{equation}
\mathbb{P} = \left( \begin{array}{ccc}
P_{\parallel} & P_a & P_b \\
P_a & P_{\perp 1} & P_c \\
P_b & P_c & P_{\perp 2} \end{array} \right)\text{,}
\end{equation}
it is always possible to find a coordinate system in which the final
two diagonal entries are equal.  To do so, rotate around the magnetic
field direction with the rotation matrix
\begin{equation}
\mathbb{R} = \left( \begin{array}{ccc} 1 & 0 & 0 \\ 0 & \cos\theta &
  -\sin\theta \\ 0 & \sin\theta & \cos\theta \end{array} \right)\text{.}
\end{equation}
Calculating $\mathbb{R}^{T}\mathbb{P}\mathbb{R}$ and setting the
second and third diagonal components equal gives the required angle:
\begin{equation}
\tan(2\theta) = \frac{P_{\perp 2} - P_{\perp 1}}{2P_c}\text{.}
\end{equation}
In the new frame the final two diagonal components are $(P_{\perp
  1}+P_{\perp 2})/2$.  The off-diagonal elements must also be known in
the new coordinate system in order to compute $Q$.  Although explicit
expressions for them can be calculated, the algebra is somewhat
tedious.

An easier method relies on the fact that certain combinations of
tensor elements are invariant under coordinate rotations.  Through
their use, $Q$ can be computed when $\mathbb{P}$ is known in any
coordinate system.  In particular, a $3\times 3$ tensor with elements
$P_{ij}$ in an arbitrary Cartesian system with coordinates $(x,y,z)$
has three invariants, two of which are the trace
\begin{equation}
I_1 = P_{xx}+P_{yy}+P_{zz}
\end{equation}
and the sum of principal minors
\begin{multline}
I_2 = P_{xx}P_{yy}+P_{xx}P_{zz}+P_{yy}P_{zz}\\-(P_{xy}P_{yx}+P_{xz}P_{zx}+P_{yz}P_{zy})\text{ .}
\end{multline}
For symmetric matrices $P_{ij}=P_{ji}$ and the final three terms of
$I_2$ simplify. The third invariant, not used here, is the
determinant.

With these definitions $Q$ can be expressed in terms of $I_1$, $I_2$,
and the parallel pressure $P_{\parallel}$.  The latter can be
calculated from 
\begin{equation}
\begin{split}
P_{\parallel} &=
\widehat{\mathbf{b}}\boldsymbol{\cdot}\mathbb{P}\boldsymbol{\cdot}\widehat{\mathbf{b}}\\ &=
b_x^2P_{xx} + b_y^2P_{yy} + b_z^2P_{zz} + \\ & \qquad
2(b_xb_yP_{xy}+b_xb_zP_{xz}+b_yb_zP_{yz})
\end{split}
\end{equation}
where $b_i$ is the $i^{\text{th}}$ component of the unit vector aligned with the
local magnetic field.  After some algebraic manipulations,
\begin{equation}\label{qcalc}
Q = 1 - \frac{4I_2}{(I_1-P_{\parallel})(I_1+3P_{\parallel})}\text{ .}
\end{equation}
With this formula $Q$ can be simply computed in an arbitrarily
oriented coordinate system, without need for coordinate
transformations and the associated matrix multiplications.



\end{document}